\documentclass[prl,twocolumn,showpacs,preprintnumbers,amsmath,amssymb,floatfix]{revtex4}

\usepackage{graphicx}   % Include figure files
\usepackage{dcolumn}    % Align table columns on decimal point
\usepackage{bm}         % bold math

\def\tthyphen{\discretionary{-}{}{-}}

\input{Qcircuit}

\begin{document}

\title{Efficient classical-communication-assisted local simulation of
n-qubit GHZ correlations}

\author{Tracey E. Tessier}

\email{tessiert@info.phys.unm.edu}

\author{Ivan H. Deutsch}

\author{Carlton M. Caves}

\affiliation{Department of Physics and Astronomy, University of New
Mexico, Albuquerque, New Mexico 87131}

\begin{abstract}
We present a local hidden-variable model supplemented by classical
communication that reproduces the quantum-mechanical predictions for
measurements of all products of Pauli operators on an $n$-qubit GHZ
state (or ``cat state'').  The simulation is efficient since the
required amount of communication scales linearly with the number of
qubits, even though there are Bell-type inequalities for these states
for which the amount of violation grows exponentially with $n$. The
structure of our model yields insight into the Gottesman-Knill theorem
by demonstrating that, at least in this limited case, the correlations
in the set of nonlocal hidden variables represented by the stabilizer
generators are captured by an appropriate set of local hidden variables
augmented by $n-2$ bits of classical communication.
\end{abstract}

\date{\today{}}


\maketitle

\section{Introduction}

Bell's theorem \cite{Bell64} codifies the observation that entangled
quantum-mechanical systems exhibit stronger correlations than are
achievable within any local hidden-variable (LHV)  model.  Beyond
philosophical implications, the ability to operate outside the
constraints imposed by local realism serves as a resource for many
information processing tasks such as communication \cite{Schumacher96}
and cryptography~\cite{Ekert91}.

The role of entanglement in quantum computation \cite{Nielsen00} is
less clear, for the issue is not one of comparing quantum predictions
to a {\it local\/} realistic description, but rather one of comparing a
quantum computation to the {\it efficiency\/} of a realistic
simulation.  Nevertheless, various results indicate some connection
between entanglement and computational
power~\cite{Steane00,Raussendorf01}. Entanglement is a necessity if a
pure-state quantum computer is to have scalable physical resources
\cite{Blume02}.  Moreover, systems with limited entanglement can often
be efficiently simulated classically~\cite{Vidal03a}. Jozsa and Linden
\cite{Jozsa02} showed that if the entanglement in a quantum computer
extends only to some fixed number of qubits, independent of problem
size, then the computation can be simulated efficiently on a classical
computer.

Despite these results, global entanglement is by no means sufficient
for achieving an exponential quantum advantage in computational
efficiency~\cite{Vidal03}.  The set of Clifford gates (Hadamard, Phase,
and CNOT) acting on a collection of $n$ qubits, each initialized to a
fiducial state $|0\rangle$, can generate globally entangled states, yet
according to the Gottesman-Knill (GK) theorem~\cite{Nielsen00}, the
outcomes of all measurements of products of Pauli operators can be
simulated with $O(n^{2})$ resources \cite{Aaronson04} on a classical
computer. The GK theorem is an expression of properties of the
$n$-qubit Pauli group ${\cal P}_n$ \cite{Nielsen00}, which consists of
all products of Pauli operators multiplied by $\pm1$ or $\pm
i\hspace{0.5pt}$: the allowed (Clifford) gates preserve ${\cal P}_n$,
and the allowed measurements are the Hermitian operators in ${\cal
P}_n$.

One approach to understanding the information processing capabilities
of entangled states is to translate a quantum protocol involving
entanglement into an equivalent protocol that utilizes only classical
resources, e.g., the shared randomness of LHVs and ordinary classical
communication.  Toner and Bacon \cite{Toner03} showed that the quantum
correlations arising from local projective measurements on a maximally
entangled state of two qubits can be simulated exactly using a LHV
model augmented by just a single bit of classical communication.
Pironio \cite{Pironio03} took this analysis a step further, showing
that the amount of violation of a Bell inequality imposes a lower bound
on the average communication needed to reproduce the quantum-mechanical
correlations.

In this article we analyze the classical resources required to simulate
measurements made on the $n$-qubit GHZ state \cite{Greenberger89} (also
called a ``cat state'').  We present a LHV model, augmented by
classical communication, that simulates the quantum-mechanical
predictions for measurements of arbitrary products of Pauli operators
on this state. The simulation is efficient since the required amount of
communication scales linearly with $n$.  These results are somewhat
surprising in light of the existence of Bell-type inequalities for
$n$-qubit GHZ states where the amount of violation grows exponentially
in the number of qubits \cite{Mermin90}.

Since the $n$-qubit GHZ state is generated by a circuit composed solely
of Clifford gates, and since we consider only measurements of
observables in ${\cal P}_n$, our result yields an alternative
perspective on the GK theorem.  Whereas the GK simulation tracks the
evolution of {\em nonlocal\/} hidden variables that specify the
generators of the stabilizer \cite{Nielsen00,Aaronson04}, we simulate
the circuit using {\em local\/} hidden variables that are supplemented
by an {\it efficient\/} amount of classical communication to predict
measurement outcomes.  We conjecture that our result is general, i.e.,
that any GK circuit can be simulated with a LHV model plus an amount of
communication that scales at most polynomially in the number of qubits.
The existence of such an efficient classical model is currently under
investigation.

\begin{figure}
\hspace{-90pt}
\Qcircuit @C=1.4em @R=1.6em {
    \lstick{\ket{0}} & \gate{H} & \ctrl{1}  & \ctrl{2}
    & \qw \\
    \lstick{\ket{0}} & \qw      & \targ     & \qw
    & \rstick{\displaystyle{\ket{\psi_{3}}=
        \frac{1}{\sqrt{2}}\left(\ket{000}+\ket{111}\right)}} \qw \\
    \lstick{\ket{0}} & \qw      & \qw       & \targ
    & \qw \\
}
\caption{Circuit to generate the three-qubit GHZ state.}
\label{fig:GHZ}
\end{figure}
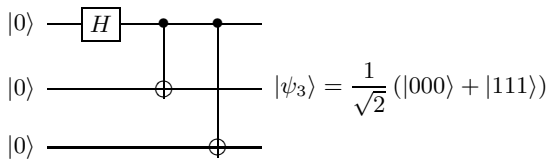

Consider now the three-qubit GHZ state, $\left|\psi_{3}\right\rangle
=\left(\left|000\right\rangle +\left|111\right\rangle
\right)/\sqrt{2}$, generated by the quantum circuit shown in
Fig.~\ref{fig:GHZ}.  In the language of the GK theorem, the evolution
of the state is tracked by the evolution of the stabilizer generators.
The Hadamard gate $H$ transforms the Pauli operators $X,Y,Z$ according
to $HXH^{\dagger}=Z$, $HYH^{\dagger}=-Y$, and $HZH^{\dagger}=X$.
Similarly, under the action of CNOT, we have
\begin{eqnarray}
&&C \left( X  I \right) C^{\dagger} = X  X\;,\quad
C \left( I  X \right) C^{\dagger} = I  X\;,\nonumber\\
&&C \left( Y  I \right) C^{\dagger} = Y  X\;,\quad
C \left( I  Y \right) C^{\dagger} = Z  Y\;,\nonumber\\
&&C \left( Z  I \right) C^{\dagger} = Z  I\;,\quad
C  \left( I  Z \right) C^{\dagger} =Z  Z\;,
\label{eq:Ctrans}
\end{eqnarray}
where the first qubit is the control, the second is the target, and $I$
represents the identity operator.  The stabilizer generators evolve
through the circuit in Fig.~\ref{fig:GHZ} as $\left\langle ZII,IZI,IIZ
\right\rangle \xrightarrow{H_{1}} \left\langle XII,IZI,IIZ\right\rangle
\xrightarrow{CNOT_{12}} \left\langle XXI, ZZI, IIZ \right\rangle
\xrightarrow{CNOT_{13}} \left\langle XXX, ZZI, ZIZ \right\rangle$.  The
full final stabilizer, consisting of all products of the generators,
includes $-XYY$, $-YXY$, $-YYX$, and $XXX$.  This means that
$|\psi_3\rangle$ is a $+1$ eigenstate of these four operators, which
implies a deterministic violation of the assumptions of local realism
\cite{Mermin90a}.

The GK description provides an efficient method for simulating the
outcome of a measurement of any product of Pauli operators on the
globally entangled state $\left|\psi_{3}\right\rangle$, but it does so
via the nonlocal stabilizer generators.  We replace this nonlocal
resource with a local description, augmented by classical
communication, by constructing a LHV table where each row represents a
qubit and each column represents a measurement.  Locality is enforced
by only allowing changes in rows corresponding to qubits that
participate in an interaction.

For the initial state $\left|000\right\rangle$, a measurement of $Z$ on
any qubit yields $+1$ with certainty, and a measurement of $X$ or $Y$
yields $\pm 1$ with equal probabilities.  The first table in
Fig.~\ref{fig:LHV} gives corresponding LHVs for this state, with
$R_{j}$ denoting a classical random variable that returns $\pm 1$ with
equal probability and $j$ labeling the qubit to which the random
variable refers.  The table is read by choosing a measurement and
multiplying the corresponding entries.  The resulting product, with $i$
discarded whenever it appears, is the outcome predicted by the LHV
model.  The LHV table yields the correct quantum-mechanical predictions
for measurements of the $4^{3}=64$ products of Pauli operators on the
state $\left|000\right\rangle $. The use of $i$ in the model,
apparently just a curiosity, actually plays a crucial role.  It
simulates some of the conflicting predictions of commuting LHVs and
anticommuting quantum operators, which are the basis of Mermin's GHZ
argument \cite{Mermin90a}. In addition, modeling the CNOT gates relies
on the $X$ and $Z$ entries being real and the $Y$ entries being
imaginary and on the particular initial correlations between the $X$
and $Y$ values for each qubit.

The first step in creating the three-qubit GHZ state is to apply the
Hadamard gate to the first qubit.  We extract rules for updating the
LHV table from the transformations of the Pauli operators, which
suggest that to simulate $H$, we should swap the $X$ and $Z$ entries,
i.e., $X^a=Z^b$ and $Z^a=X^b$, and flip the sign of the $Y$ entry,
i.e., $Y^a=-Y^b$, where $b$ and $a$ denote LHV values before and after
a gate.  Applying these rules to the first row leads to the second
table in Fig.~\ref{fig:LHV}, which returns correct quantum-mechanical
predictions for all measurements of Pauli products on the state
$\left(\left|0\right\rangle+\left|1\right\rangle\right)
\left|00\right\rangle/\sqrt2$.  This is not surprising since the state
remains a product state, and it is well known that a LHV model can be
constructed for a single qubit \cite{Bell64}.  The usefulness of our
model only becomes apparent when we apply it to entangled states.

\begin{figure}
$
\begin{array}{cccc}
X & Y & Z \\
R_{1} & -iR_{1} & 1 \\
R_{2} & iR_{2} & 1 \\
R_{3} & iR_{3} & 1 \end{array}
\xrightarrow{H_{1}}
\begin{array}{cccc}
X & Y & Z\\
1 & iR_{1} & R_{1}\\
 R_{2} & iR_{2} & 1\\
 R_{3} & iR_{3} & 1
\end{array}
\xrightarrow{\mbox{\scriptsize{CNOT}$_{12}$}}
\begin{array}{cccc}
 X & Y & Z\\
R_{2} & iR_{1}R_2 & R_{1}\\
R_{2} & iR_{1}R_2 & R_{1}\\
R_{3} & iR_{3} & 1
\end{array}
\vspace{6pt}
\newline
\xrightarrow{\mbox{\scriptsize{CNOT}$_{13}$}}
\begin{array}{cccc}
X & Y & Z\\
R_{2}R_3 & iR_{1}R_2R_3 & R_{1}\\
R_{2} & iR_{1}R_2 & R_{1}\\
R_{3} & iR_{1}R_3 & R_{1}
\end{array}
$ \caption{Evolution of the LHV model during creation of the
three-qubit GHZ state.  The rules for updating the LHV tables are
suggested by the equations for transforming Pauli
operators.\label{fig:LHV}}
\end{figure}

Applying the first CNOT gate in Fig.~\ref{fig:GHZ} yields the Bell
entangled state $\left(\left|00\right\rangle+
\left|11\right\rangle\right)\left|0\right\rangle/\sqrt2$.  To update
the LHV table entries under a CNOT, we use the following rules for the
control~$c$ and the target~$t$:
\begin{eqnarray}
&&X_c^a = X_c^b X_t^b\;,\quad
Y_c^a = Y_c^b X_t^b \;,\quad
Z_c^a = Z_c^b  \;,\nonumber\\
&&X_t^a = X_t^b\;,\quad
Y_t^a = Z_c^b Y_t^b \;,\quad
Z_t^a = Z_c^bZ_t^b
\;.
\label{eq:Crules}
\end{eqnarray}
The update rules for $H$ and CNOT keep the $X$ and $Z$ entries real and
the $Y$ entry imaginary, and the CNOT rule preserves the correlation
$XYZ=i$ that holds for each qubit after the operation of the Hadamard
gate.  Using the rules~(\ref{eq:Crules}) in the first two rows gives
the third table in Fig.~\ref{fig:LHV} to represent the Bell state.

The LHV rules~(\ref{eq:Crules}) must be consistent with the fifteen
transformations of nontrivial Pauli products under CNOT.  For example,
the transformation $C\left(XI\right)C^\dagger=XX$ requires that
$X_c^b=X_c^aX_t^a$, which does follow from the rules~(\ref{eq:Crules}).
The CNOT rules~(\ref{eq:Crules}) are derived from the six
transformations~(\ref{eq:Ctrans}), and because $C=C^\dagger$, these
rules are consistent with five other transformations. Consistency with
the remaining four transformations, those being $C \left( X Y \right)
C^{\dagger} = Y Z$, $C \left( X Z \right) C^{\dagger} = - Y Y$, and the
inverse transformations, requires that
$X_c^bY_t^b=Y_c^aZ_t^a=Y_c^bZ_c^bZ_t^bX_t^b$ and
$X_c^bZ_t^b=-Y_c^aY_t^a=-Y_c^bZ_c^bX_t^bY_t^b$.  These relations do not
hold generally, but they are satisfied if the initial entries for both
the control and target are correlated according to $XYZ=i$, with $X$
and $Z$ real and $Y$ imaginary.  In all our applications of CNOT, these
conditions hold.   That they are not generally true is the chief
obstacle to extending our results to arbitrary GK circuits and the
entangled states they produce.

The third table in Fig.~\ref{fig:LHV} gives the correct
quantum-mechanical predictions for all measurements of Pauli products
on the Bell state
$\left(\left|00\right\rangle+\left|11\right\rangle\right)
\left|0\right\rangle/\sqrt2$.  What is new are the correlations between
the first two qubits in each column.  For example, the single-qubit
measurements $ZII$ and $IZI$ both return the random result $R_1$; the
product of these outcomes always equals $+1$, the same as the outcome
of a joint measurement of $ZZI$ on the first two qubits. In this
context, the $i$'s in the correlated $Y$ entries now lead to a problem:
the single-qubit measurements $YII$ and $IYI$ both give the random
result $R_1R_2$, with product $+1$, inconsistent with the outcome
$(iR_1R_2)(iR_1R_2)=-1$ of a joint measurement of $YYI$ \cite{problem}.
This problem persists throughout our analysis, occurring for joint
measurements involving $Y$'s on some qubits and having outcomes that
are certain (i.e., measurements of stabilizer elements).  This is the
reason our LHV model must be supplemented by classical communication.

At this point the problem is restricted to the joint measurements $YYI$
and $YYZ$ and the corresponding local measurements and thus can be
corrected by flipping the sign of the outcome whenever a local
measurement of $Y$ is made on the first qubit; i.e., the model returns
the random result $-R_1R_2$ for a measurement of $YII$.  This sign flip
fixes the required correlations and is irrelevant to other joint
measurements that involve $Y$ on the first qubit, all of which have
random results.  Since the sign flip depends only on the measurement on
the first qubit, it requires {\em no\/} communication between the
qubits.  Thus at this stage, with Bell-state entanglement, the LHV
model gives correct quantum-mechanical predictions for all observables
in ${\cal P}_3$ and their correlations.

We complete the simulation of the creation of the GHZ state by
performing the CNOT between the first and third qubits, resulting in
the last table in Fig.~\ref{fig:LHV}.  This table yields correct
quantum-mechanical predictions for all of the observables in ${\cal
P}_{3}$, including those that form the basis of Mermin's GHZ argument
\cite{Mermin90a}, i.e., $XXX=1$ and $XYY=YXY=YYX=-1$. As promised, the
imaginary $Y$ entries make this agreement possible.

Consider now the scheme for ensuring consistency with local measurement
predictions for the three-qubit GHZ state.  The only local measurements
that yield inconsistent results are those associated with stabilizer
elements that contain $Y$'s, i.e., the joint measurements $XYY$, $YXY$,
and $YYX$.  Let Alice, Bob, and Carol each possess one of the qubits.
If we put Alice in charge of ensuring compatibility, she should flip
the sign of her outcome whenever she and/or Bob measures $Y$ locally.
This sign flip fixes the local correlations associated with $XYY$,
$YXY$, and $YYX$ and is irrelevant to other possible joint measurements
that involve $Y$'s on the first two qubits, all of which have random
outcomes.  To implement this scheme, Bob must communicate to Alice one
bit denoting whether or not he measured $Y$.  For the three-qubit GHZ
state, we thus have a LHV model, assisted by one bit of classical
communication, that duplicates the quantum-mechanical predictions for
all measurements in ${\cal P}_{3}$ and their correlations.

The circuit that creates the general $n$-qubit GHZ state,
$\left|\psi_{n}\right\rangle =\left(\left|00\ldots0
\right\rangle\*+\left|11\ldots1\right\rangle \right)/\sqrt2$, has the
same topology as in Fig.~\ref{fig:GHZ}: a Hadamard on the first qubit
is followed by $n-1$ CNOT gates, with the leading qubit as the control
and the remaining qubits serving successively as targets.  The operator
transformations show that $|\psi_n\rangle$ is specified by the $n$
stabilizer generators $\langle X^{\otimes n}, ZZI^{\otimes (n-2)},
ZIZI^{\otimes (n-3)},\ldots, ZI^{\otimes (n-2)} Z\rangle$.  The full
stabilizer consists of the $2^n$ observables in ${\cal P}_n$ that yield
$+1$ with certainty.  It contains Pauli products that have (i)~only
$I$'s and an even number of $Z$'s and (ii)~only $X$'s and an even
number of $Y$'s, with an overall minus sign if the number of $Y$'s is
not a multiple of 4. Of the remaining $2\times 4^n$ observables in
${\cal P}_n$, $2^n$ are negatives of the stabilizer elements, thus
yielding $-1$, while the rest return $\pm 1$ with equal
probability~\cite{Nielsen00}.

Following the same procedure as in the three-qubit case, one finds that
the LHV table representing the $n$-qubit GHZ state is given by
\begin{equation}
\begin{array}{cccc}
 & X & Y & Z\\
\mbox{qubit 1}  & R_2R_3\cdots R_n & iR_1R_2\cdots R_n & R_{1}\\
\mbox{qubit 2}  & R_{2} & iR_{1}R_2 & R_{1}\\
\mbox{qubit 3}  & R_{3} & iR_{1}R_3 & R_{1}\\
\vdots & \vdots & \vdots & \vdots\\
\mbox{qubit $n$}& R_{n} & iR_{1}R_n & R_{1}
\end{array}\;\;.\label{eq:NQubitGHZLHV}
\end{equation}
That this table gives the correct quantum-mechanical predictions for
all measurements of Pauli products follows from the consistency of our
LHV update rules, but it is nevertheless useful to check this directly.
Suppose a Pauli product contains no $X$'s or $Y$'s, but consists solely
of $I$'s and $Z$'s.  Then it is clear from the table in
Eq.~(\ref{eq:NQubitGHZLHV}) that the outcome is certain if and only if
the number of $Z$'s in the product is even.  Suppose now that the
product has an $X$ or a $Y$ in the first position.  Then it is apparent
that to avoid a random variable in the overall product, all the other
elements in the product must be $X$'s or $Y$'s and the number of $Y$'s
must be even; the outcome is $+1$ if the number of $Y$'s is a multiple
of 4 and $-1$ otherwise.  Finally, suppose the Pauli product has an $X$
or a $Y$ in a position other than the first.  Then the only way to
avoid a random variable in the overall product is to have an $X$ or a
$Y$ in the first position, and we proceed as before. This argument
shows that the LHV table for the $n$-qubit GHZ state gives correct
quantum-mechanical predictions for measurements of all Pauli products.

It remains to ensure that the products of the LHV predictions for local
measurements are consistent with the corresponding joint measurement
results.  As before, the source of the inconsistency is the
$i\hspace{0.5pt}$ in the $Y$ table entries, the very thing that allows
us to get all the Pauli products correct.  Stationing Alice at the
first qubit and putting her in charge of ensuring consistency, we see
that what she needs to know is the number of $i$'s in the product for
the corresponding joint measurement.  In particular, letting $q_j=i$ if
$Y$ is measured on the $j$th qubit and $q_j=1$ otherwise, Alice can
ensure consistency by changing the sign of her local outcome if the
product $p_n=q_1\cdots q_n$ is $-1$ or $-i$ and leaving her local
outcome unchanged if $p_n$ is $+1$ or $i$.  This scheme requires $n-1$
bits of communication as each of the other parties communicates to
Alice whether or not he measured $Y$, but we can do a bit better.
Alice's action is only important when $p_n$ is $+1$ or $-1$; when $p_n$
is $i$ or $-i$, the sign flip or lack thereof is irrelevant because the
joint measurement outcome is random. As a result, Alice can get by with
the truncated product $p_{n-1}=q_1\cdots q_{n-1}$: she flips the sign
of her local outcome if $p_{n-1}$ is $i$ or $-1$ and leaves the local
outcome unchanged if $p_{n-1}$ is $-i$ or $1$.  The scheme works
because whether $q_n$ is 1 or $i$, Alice flips when $p_n=-1$ and
doesn't flip when $p_n=+1$, as required.  This improved scheme requires
$n-2$ bits of classical communication; it generalizes our previous
results for the Bell state and the three-qubit GHZ state.

The consistency scheme generalizes trivially to the case of
measurements made on $l$ disjoint sets of qubits.  For each set $k$
chosen from the $l$ sets, the table yields a measurement product that
is the predicted outcome multiplied by $q_k=i$ or $q_k=1$. Letting
Alice be in charge of the first set, all but the last of the other sets
communicates $q_k$ to Alice, who computes the product $q_1\cdots
q_{l-1}$ and decides whether to flip her set's outcome just as before.
Consistency with the corresponding joint measurement is thus ensured at
the price of $l-2$ bits of communication.

Using local hidden variables and an efficient amount of classical
communication, we have shown that it is possible to simulate the
correlations that arise when measuring arbitrary products of Pauli
operators on an $n$-qubit GHZ state.  Though the $n$-qubit GHZ state is
highly entangled, the probability distributions for the allowed
measurements of Pauli products are essentially trivial, being either
certainty or binary randomness.  This property is shared by all states
produced by GK circuits, leading us to conjecture that our results can
be extended to measurements of Pauli products on any state produced by
a GK circuit.  In contrast, allowing just one additional nontrivial
measurement, say of $\left(X+Z\right)/\sqrt2$, leads to correlations
for which our simple simulation will no longer work.  We anticipate
that under this more general measurement scheme, the amount of
classical communication required to make a LHV model work grows
exponentially in the number of qubits.

Our model provides weak evidence that the power of quantum computation
arises not directly from entanglement, but rather from the lack of an
efficient, local realistic description assisted by an efficient amount
of nonlocal, but classical communication.  An efficient
communication-assisted LHV model for all GK circuits would provide
powerful additional evidence for this idea.

\begin{acknowledgments}
We thank D.~Bacon and R.~Raussendorf for helpful discussions. The
quantum circuit in Fig.~\ref{fig:GHZ} was set using the \LaTeX\ package
{\tt Qcircuit}, available at http://info.phys.unm.edu/Qcircuit/. This
work was partly supported by ARO Grant No.~DAAD19-01-1-0648.
\end{acknowledgments}


\begin{thebibliography}{99}
\bibitem{problem}
This problem can be transformed back to the initial state: a
measurement of $YXI$ has result $-R_1R_2$, and a measurement of $XYI$
has the opposite result, $R_1R_2$, for a product of $-1$, whereas a
measurement of $ZZI$ has result $+1$ not equal to the product.  This is
not seen as a problem because there is no connection between these
measurements in a LHV model.
\end{thebibliography}
\end{document}